\documentclass[aps,pre,groupedaddress,twocolumn,10pt]{revtex4-1}
\usepackage{graphicx}
\usepackage[english]{babel}
\usepackage[T1]{fontenc}

\usepackage{amsmath}
\usepackage{amssymb}
\usepackage{amsthm}
\usepackage{bbm}
\usepackage{hyperref}
\usepackage[usenames,dvipsnames]{xcolor}
\hypersetup{colorlinks=true, linkcolor=BrickRed, urlcolor=blue!50!black, citecolor=blue!50!black}

\usepackage[normalem]{ulem}

\newcommand\diff{\mathrm{d}}
\renewcommand{\vec}[1]{\mathbf{#1}}
\renewcommand{\imath}[0]{\mathsf{i}}

\definecolor{cream}{RGB}{222,217,201}

\usepackage[nodayofweek,level]{datetime}

\begin{document}

\title{Elastic behavior of a semiflexible polymer in 3D subject to compression and stretching forces}

\author{Christina Kurzthaler}
\affiliation{Institut f\"ur Theoretische Physik, Universit\"at Innsbruck, Technikerstra{\ss}e 21A,
A-6020 Innsbruck, Austria}
\email{christina.kurzthaler@uibk.ac.at}

\begin{abstract}
We elucidate the elastic behavior of a wormlike chain in 3D under compression and provide exact solutions for the experimentally accessible force-extension relation in terms of generalized spheroidal wave functions. In striking contrast to the classical Euler buckling instability, the force-extension relation of a clamped semiflexible polymer exhibits a smooth crossover from an almost stretched to a buckled configuration. In particular, the associated susceptibility, which measures the strength of the response of the polymer to the applied force, displays a prominent peak in the vicinity of the critical Euler buckling force. For increasing persistence length, the force-extension relation and the susceptibility of semiflexible polymers approach the behavior of a classical rod, whereas thermal fluctuations permit more flexible polymers to resist the applied force. Furthermore, we find that semiflexible polymers confined to 2D can oppose the applied force more strongly than in 3D.
\end{abstract}

\date{\formatdate{21}{8}{2018}}

\maketitle

Semiflexible polymers are ubiquitous in nature~\cite{MacKintosh:2014} and include filamentous actin~\cite{Lieleg:2010}, microtubules~\cite{Brangwynne:2006}, intermediate filaments~\cite{Nolting:2014}, and DNA strands~\cite{Marko:1995} as most prominent examples. These biopolymers represent integral parts of cells and are crucial for their elastic properties and shape, intracellular transport processes, and mitosis~\cite{Brangwynne:2006, Fletcher:2010,Lieleg:2010}. In particular, they are often connected by regulatory proteins and form complex networks, which exhibit peculiar elastic behavior induced by their confining surroundings~\cite{Bausch:2006}. Semiflexible polymers also display intriguing non-equilibrium transport properties in the presence of motor proteins evoking self-assembly and pattern formation~\cite{Schaller:2010,Shelley:2016}. Understanding this intricate many-body physics relies on a profound knowledge of the underlying physical mechanisms at different levels of coarse-graining.

The elastic properties of single semiflexible polymers have been elucidated for purified DNA~\cite{Bouchiat:1999,Bustamante:2000,Marko:1995,Sitters:2015} and actin filaments~\cite{Liu:2002}, single molecules, as, for example, titin~\cite{Kellermayer:1997} and collagen~\cite{Sun:2002}, and also synthetic carbon nanotubes~\cite{Kuzumaki:2006} in terms of force-extension relations. These have been measured by using optical~\cite{Ashkin:1997,Mehta:1999} and magnetic tweezers~\cite{Gosse:2002}, acoustic~\cite{Sitters:2015} and atomic force spectroscopy~\cite{Hugel:2001,Janshoff:2000}, or transmission electron microscopy~\cite{Kuzumaki:2006} and revealed a strikingly nonlinear response to extension and compression forces. In particular, the behavior of a semiflexible polymer is drastically altered by thermal fluctuations and, in contrast to the classical Euler buckling instability of a rigid rod~\cite{Landau:1986}, it is determined by the interplay of enthalpic elasticity and conformational entropy~\cite{MacKintosh:2014}.

To characterize the elasticity of a semiflexibe polymer, the celebrated wormlike chain model, also referred to as Kratky-Porod model~\cite{Kratky:1949}, has been employed and analyzed in terms of exact low-order moments for the end-to-end distance~\cite{Hamprecht:2004, Spakowitz:2004, Spakowitz:2005, Mehraeen:2008}, numerical solutions for the associated probability distribution~\cite{Samuel:2002, Mehraeen:2008}, and the probability densities of polymers in the weakly-bending regime~\cite{Wilhelm:1996}. Exact solutions for the force-extension relation of a wormlike chain in 2D have been provided for stretching~\cite{Prasad:2005} and only recently for compression forces~\cite{Kurzthaler:2017}, which complete previous approximate theories in the regime of stiff polymers~\cite{Baczynski:2007,Emanuel:2007,Lee:2007,Bedi:2015}. Furthermore, the response of a wormlike chain in 2D has been elucidated using exact solutions for the associated probability densities~\cite{Kurzthaler:2018}.
For the 3D case, the force-extension relation of a semiflexible polymer under compression has been elaborated in Ref.~\cite{Pilyugina:2017}, where the stretching energy depends on the end-to-end distance of the polymer by keeping its ends free. In particular, this guarantees that the force is always compressive. Here, the Green function of a semiflexible polymer has been elaborated exactly in terms of a continued fraction representation, which permitted to extract the exact force-extension relation of a semiflexible polymer under compression from the associated minimal Helmholtz free energy. Configuration snapshots have been obtained via Monte Carlo simulations.

Here, we study the elastic behavior of a semiflexible polymer subject to a constant external force along a fixed direction, which constitutes the relevant scenario in most stretching experiments. We provide exact solutions for the force-extension relation of a wormlike chain in 3D under compression. We consider different bending rigidities and account for clamped, cantilevered (i.e. one clamped and one free end), and free ends of the polymer reflecting different experimental setups. Moreover, we compute the associated susceptibilities to characterize the strength of the response of the polymer to the applied compression force. We also  compare our results for a polymer in 3D to the behavior of a polymer in 2D and corroborate our results by stochastic pseudo-dynamics simulations.

\section{Wormlike chain model\label{sec:model}}
To describe the elastic behavior of a semiflexible polymer we rely on the well-established wormlike chain model~\cite{Kratky:1949}. The bending energy of an inextensible  wormlike chain is expressed by its integrated squared curvature,
\begin{align}
  \mathcal{H}_0	&= \frac{\kappa}{2}\int_0^L\diff s \ \left(\frac{\diff \vec{u}(s)}{\diff s}\right)^2,
\end{align}
where $s$ is the arc length, $L$ the contour length, and  $\kappa$ the bending stiffness of the polymer. The tangent vector of the polymer along its contour $\vec{r}(s)$ is denoted by $\vec{u}(s) = \diff \vec{r}(s)/ \diff s$ and has unit length, $|\vec{u}(s)|=1$. Thus, the partition sum of a wormlike chain with given initial and final orientations, $\vec{u}_0$ and $\vec{u}_L$, can be regarded as sum over all possible chain configurations, which obey the inextensibility constraint $|\vec{u}(s)|=1$, weighted by the Boltzmann factor. Formally, it is obtained as a path integral
\begin{align}
  Z_0(\vec{u}_L,L|\vec{u}_0,0)&\!=\! \int _{\vec{u}(0)=\vec{u}_0}^{\vec{u}(L)=\vec{u}_L}\!\mathcal{D}[\vec{u}(s)]\exp\left(-\frac{\mathcal{H}_0}{k_\text{B}T}\right),
\end{align}
where $k_\text{B} T$ denotes the thermal energy with Boltzmann constant $k_\text{B}$ and temperature $T$. Due to the local inextensibility constraint, the path integral cannot be calculated by Gaussian integrals, but exact solutions can be elaborated by solving the corresponding Fokker-Planck equation~\cite{Spakowitz:2004}. The solutions permit to introduce the persistence length, $\ell_\text{p}=\kappa/k_\text{B}T$ for 3D and $\ell_\text{p}=2\kappa/k_\text{B}T$ for 2D, as the decay length of the tangent-tangent correlations of the polymer, $\langle\vec{u}(s)\cdot\vec{u}(s')\rangle = \exp[-(s-s')/\ell_\text{p}]$. The persistence length represents a geometric measure for the stiffness of the polymer and allows classifying polymers into flexible $\ell_\text{p}/L \ll 1$, semiflexible $\ell_\text{p}/L \simeq 1$, and stiff polymers $\ell_\text{p}/L \gg 1$~\cite{Doi:1986, MacKintosh:2014}.

\begin{figure}[htp]
\centering
\includegraphics[width = \linewidth]{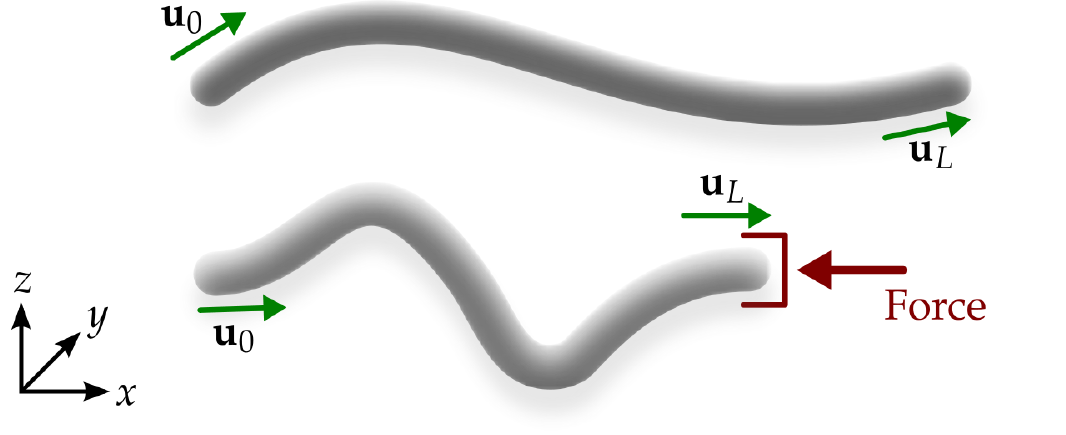}
\caption{Sketch of a semiflexible polymer in 3D with initial orientation $\vec{u}_0$ and final orientation $\vec{u}_L$
under compression.\label{fig:sketch}}
\end{figure}

Moreover, we consider the wormlike chain subject to a constant external force $\vec{F}$ (see Fig.~\ref{fig:sketch}). Thus, the stretching energy
\begin{align}
  \mathcal{H}_\text{force} &= -\vec{F} \cdot\left[\vec{r}(L)-\vec{r}(0)\right],\label{eq:H_force}
\end{align}
contributes to the full energy of the system. Here, the force $\vec{F}=F\vec{e}$ acts along a fixed direction $\vec{e}$ of unit length $|\vec{e}|=1$ with strength $F$, which can be either tensile $F>0$ or compressive $F<0$. Collecting terms, the full Hamiltonian of the system assumes the form
\begin{align}
  \frac{\mathcal{H}}{k_\text{B} T}
  &= \int_0^L \diff s \ \left[\frac{\kappa}{2k_\text{B}T}  \left(\frac{\diff \vec{u}(s)}{\diff s}\right)^2 -f \vec{e}\cdot \vec{u}(s)\right], \label{eq:hamiltonian}
\end{align}
where we have introduced the reduced force $f = F/k_\text{B}T$ with units of an inverse length. The corresponding partition sum reads
\begin{align}
  Z(\vec{u}_L,L|\vec{u}_0,0)  &= \int _{\vec{u}(0)=\vec{u}_0}^{\vec{u}(L)=\vec{u}_L}\mathcal{D}[\vec{u}(s)]\exp\left(-\frac{\mathcal{H}}{k_\text{B}T}\right),\label{eq:Z}
\end{align}
and the associated thermodynamic potential represents the Gibbs free energy defined by
\begin{align}
G(T,F) &= -k_\text{B}T\ln[Z(\vec{u}_L,L|\vec{u}_0,0)]. \label{eq:gibbs}
\end{align}
It obeys the fundamental relation
\begin{align}
 \diff G &= -S\diff T -\langle X \rangle \diff F,
\end{align}
where $S=S(T,F)$ denotes the entropy of the polymer and $\langle X\rangle :=\int_0^L\diff s \ \langle \vec{e}\cdot\vec{u}(s)\rangle$ the mean end-to-end distance projected along the direction of the force $\vec{e}$. Thus, the Gibbs free energy encodes the elastic properties of a semiflexible polymer, in particular, the force-extension relation is obtained as a derivative of the Gibbs free energy with respect to the applied force,
\begin{align}
  \langle X \rangle   &= -\left(\frac{\partial G}{\partial F}\right)_T =-\frac{1}{k_\text{B}T}\left(\frac{\partial G}{\partial f}\right)_T.\label{eq:X}
\end{align}
Moreover,  the associated isothermal susceptibility or compliance,
\begin{align}
  \chi &= \left(\frac{\partial \langle X \rangle}{\partial F}\right)_T=\frac{1}{k_\text{B}T}\left(\frac{\partial \langle X \rangle}{\partial f}\right)_T,\label{eq:chi}
\end{align}
measures the strength of the response with respect to the applied force and permits to characterize the
\textit{buckling transition} of a semiflexible polymer reminiscent of the Euler buckling instability of a classical rigid rod~\cite{Landau:1986}.

\subsection{Analytic solution for the partition sum}
An analytic solution for the partition sum serves as input to further elucidate the elastic properties
of a semiflexible polymer. By standard methods~\cite{kleinert:2009} one derives the associated Fokker-Planck equation
\begin{align}
  \partial_s Z(\vec{u},s|\vec{u}_0,0)  	&=\left[f\vec{e}\cdot\vec{u}+ \frac{k_\text{B}T}{2\kappa} \Delta_{\vec{u}}\right]Z(\vec{u},s|\vec{u}_0,0),
\end{align}
which describes the evolution of the partition sum of a semiflexible polymer along the arc length $s$ given the initial orientation $\vec{u}_0$ at $s=0$. Here, $\Delta_\vec{u}$ denotes the angular part of the Laplacian and the Fokker-Planck equation obeys the initial condition
\begin{align}
  Z(\vec{u},s=0|\vec{u}_0,0) &= \delta(\vec{u},\vec{u}_0),
\end{align}
such that the delta function $\delta(\cdot,\cdot)$ enforces both orientations, $\vec{u}$ and $\vec{u}_0$, to coincide. The Fokker-Planck equation is reminiscent of the Schr\"odinger equation of a quantum pendulum, which has been solved analytically in 2D in terms of Mathieu functions~\cite{Aldrovandi:1980}. Employing this analogy, exact solutions for the partition sums of a semiflexible polymer in 2D subject to tension~\cite{Prasad:2005} and compression forces~\cite{Kurzthaler:2017} have been calculated. Here, we consider a polymer in 3D and parametrize the orientation $\vec{u}$ in terms of the polar angle measured
with respect to the applied force $\vartheta=\angle(\vec{e},\vec{u})$ and the azimuth angle $\varphi$. Thus, the Fokker-Planck equation for the partition sum $Z\equiv Z(\vartheta,\varphi,s|\vartheta_0,\varphi_0,0)$ of a semiflexible polymer under compression, $f = -|f|$,
\begin{align}
 \partial_s Z &\!=\!\left[\!-|f|\cos\vartheta\!+\! \frac{1}{2\ell_\text{p}}\!\left(\frac{1}{\sin\vartheta}\partial_\vartheta\left(\sin\vartheta\partial_\vartheta\right)+
\frac{1}{\sin^2\vartheta}\partial^2_\varphi\right)\!\right]Z, \label{eq_fokker_planck_3d}
\end{align}
can be solved analytically by separation of variables in terms of appropriate angular eigenfunctions. Note, that in 2D the angular part of the Laplacian is replaced by the second derivative with respect to the polar angle~\cite{Prasad:2005,Kurzthaler:2017}.

Interestingly, the space curve of a semiflexible polymer can be formally regarded as the trajectory of a self-propelled agent and the solution strategies for both problems are intimately related to one another. In particular, we can transfer the methods developed for an active Brownian particle in 3D~\cite{Kurzthaler:2016} to derive the partition sum of a polymer in 3D. Hence, we employ the separation ansatz $\exp(-\lambda s)\exp(\imath m\varphi)w(\eta)$, where we have abbreviated $\eta = \cos\vartheta$,
and obtain the eigenvalue problem
\begin{align}
&\left[\!-|f|\eta+\frac{1}{2\ell_\text{p}}\left( \frac{\diff}{\diff \eta}\left[\left(1-\eta^2\right)\frac{\diff}{\diff \eta}\right]-\frac{m^2}{1-\eta^2}\right)\right]w(\eta)=\notag\\
& \qquad \qquad \qquad \qquad \qquad \qquad = -\lambda w(\eta).
\end{align}
Rearranging terms, we relate the eigenvalue problem to the generalized spheroidal wave equation~\cite{NIST:DLMF,Yan:2009, Kurzthaler:2016} with eigenfunctions $\text{Ps}_\ell^m(R,\eta)$ and eigenvalues $A_\ell^m(R)$,
\begin{align}
 L(R,\eta)\text{Ps}_\ell^m(R,\eta)&= -A_\ell^m(R)\text{Ps}_\ell^m(R,\eta),\label{eq:spheroidal_EV}
\end{align}
and corresponding Sturm-Liouville operator
\begin{align}
 L(R,\eta) &=  \frac{\diff}{\diff \eta}\left[\left(1-\eta^2\right)\frac{\diff}{\diff \eta}\right]+ R\eta - \frac{m^2}{1-\eta^2}, \label{eq:sturm_liouville}
\end{align}
dependent on the deformation parameter $R=-2\ell_\text{p}|f|$. The eigenvalues are connected to the separation constant via $A_\ell^m(R) = 2\ell_\text{p}\lambda_\ell^m$. By comparison, the eigenvalue problem of the 2D analog assumes the form of the Mathieu equation~\cite{Prasad:2005,Kurzthaler:2017}.

For $R=0$ the eigenvalue problem [Eq.~\eqref{eq:spheroidal_EV}] reduces to the general Legendre equation, where the eigenfunctions are the associated Legendre polynomials,
$\text{Ps}_\ell^m(0,\eta)\equiv|\ell,m\rangle := \sqrt{(2\ell+1)/2}\sqrt{(\ell-m)!/(\ell+m)!} \ P_\ell^m(\eta)$ of degree $\ell\in \mathbb{N}_0$ and order $m$ with corresponding eigenvalues $A_\ell^m(0)=\ell(\ell+1)$~\cite{Arfken:2005}. Thus, the generalized spheroidal wave functions can be expressed as an expansion in associated Legendre polynomials
\begin{align}
\text{Ps}_\ell^m(R,\eta) &= \sum_{j=|m|}^\infty d_j^{m\ell}(R)|j,m\rangle, \label{eq:Pslm}
\end{align}
with coefficients $d_j^{m\ell}(R)$, which are determined by the recurrence relations, Eq.~\eqref{eq:recurrence_di} in the appendix~\ref{appendix_numerics}. The generalized spheroidal wave functions represent an orthonormalized set of eigenfunctions~\cite{NIST:DLMF, Yan:2009, Kurzthaler:2016} with $\int_{-1}^1\! \diff\eta \ \text{Ps}_{n}^{m}(R, \eta)\text{Ps}_{\ell}^m(R,\eta)  = \delta_{n\ell}$.
Note, that the normalization has been employed differently in Ref.~\cite{NIST:DLMF}.

Then the full solution of Eq.~\eqref{eq_fokker_planck_3d} can be expressed as an expansion in generalized spheroidal wave functions,
\begin{align}
Z(\vec{u}_L,L|\vec{u}_0,0) = \frac{1}{2\pi}&\sum_{\ell=0}^\infty\sum_{m=-\infty}^{\infty} e^{-A_\ell^{m}(R) L/2\ell_\text{p}}e^{\imath m(\varphi_L-\varphi_0)}\notag\\
& \times \text{Ps}_\ell^{m}(R,\eta_L)\text{Ps}_\ell^{m}(R,\eta_0).\label{eq:SumZ}
\end{align}
In particular, in the partition sum of a semiflexible polymer in 2D the angular eigenfunctions are replaced by the Mathieu functions~\cite{NIST:DLMF,Kurzthaler:2017}.

\subsection{Clamped, cantilevered, and free semiflexible polymers}
We account for different experimental setups, including polymers with clamped, cantilevered, and free ends. The presented partition sum [Eq.~\eqref{eq:SumZ}] represents the case of a polymer with given initial ($\varphi_0$, $\eta_0$) and final orientations ($\varphi_L$, $\eta_L$). In our subsequent analysis we consider a clamped polymer with equal initial and final orientations, $\eta_0=\eta_L$. Furthermore, the force is applied along the direction of the initial and final orientations of the clamped polymer. Then the elastic properties are axially symmetric and we can integrate over the azimuth angles, $\varphi_0$ and $\varphi_L$. Thus, we find that only the zeroth modes contribute to the partition sum of a clamped polymer:
\begin{align}
\begin{split}
Z(\eta_L,L|\eta_0,0) &= 2\pi\sum_{\ell=0}^\infty e^{-A_\ell^{0}(R)L/2\ell_\text{p}}\text{Ps}_\ell^{0}(R,\eta_L)\text{Ps}_\ell^{0}(R,\eta_0).\label{eq:SumZ_clamped}
\end{split}
\end{align}
Interestingly, the Sturm-Liouville operator [Eq.~\eqref{eq:sturm_liouville}] for a polymer under extension, $R\mapsto -R$, obeys the symmetry relation $L(-R,\eta)=L(R,-\eta)$. Thus, the partition sum of a polymer under tension can be obtained by mapping $\eta_0 \mapsto -\eta_0$ and $\eta_L \mapsto -\eta_L$ in Eq.~\eqref{eq:SumZ_clamped}.

In the case of a cantilevered polymer the initial orientation is clamped and the final one is free. Here, the force acts along the initial orientation. The corresponding partition sum $Z(L|\eta_0,0)$ is obtained via integration over all final orientations $\eta_L$,
\begin{align}
  &Z(L|\eta_0,0)  = \int_{-1}^1\!\diff\eta_L \ Z(\eta_L,L|\eta_0,0),\notag\\
  &=  2\pi \sum_{\ell=0}^{\infty}e^{-A_\ell^0(R) L/2\ell_\text{p}}\left[\int_{-1}^1\!\diff\eta_L \ \text{Ps}_\ell^0(R,\eta_L)\right]\text{Ps}_\ell^0(R,\eta_0),\notag\\
 &=  2\sqrt{2}\pi \sum_{\ell=0}^{\infty}e^{-A_\ell^0(R) L/2\ell_\text{p}}d_0^{0\ell}(R)\text{Ps}_\ell^0(R,\eta_0)\label{eq:SumZ2},
\end{align}
where $d_0^{0\ell}(R)$ is the zeroth coefficient of the generalized spheroidal wave functions. It results
from the integral over the eigenfunction [Eq.~\eqref{eq:int_Ps} in the appendix~\ref{appendix_numerics}].

Integrating over the initial orientations $\eta_0$, the partition sum corresponding to a free polymer reduces to
\begin{align}
Z(L)  &= 4\pi\sum_{\ell=0}^{\infty}\left[d_0^{0\ell}(R)\right]^2\exp\left[-A_{\ell}^0(R)L/2\ell_\text{p}\right].\label{eq:SumZ3}
\end{align}
Due to the symmetry of the Sturm-Liouville operator [Eq.~\eqref{eq:sturm_liouville}], the partition sum for a free polymer under compression is identical to that under extension and the free ends permit the polymer to align with the direction of the applied force.

For details on the numerical evaluation of the generalized spheroidal wave functions $\text{Ps}_\ell^0(R,\eta)$ and the associated
eigenvalues $A_\ell^0(R)$ we refer to appendix~\ref{appendix_numerics}.

\subsection{Euler buckling instability}
We compare the results of a semiflexible polymer under compression to the Euler buckling instability of a classical rigid rod, which is the phenomenon that a rod under compression does not yield at all at small compression forces, but starts to bend at a critical force~\cite{Landau:1986},
\begin{align}
 F_c &= \frac{\pi^2\kappa}{(\gamma L)^2}.
\end{align}
The critical force, also referred to as critical Euler buckling force,
is determined by the bending rigidity $\kappa$, the length $L$ of the rod, and $\gamma$, which accounts for the boundary conditions. In particular, for a rod with clamped ends $\gamma=1$ and for a cantilevered rod $\gamma = 2$. Moreover, $F_c$ serves as a reference scale for the compression force applied to the semiflexible polymer and, thus, allows discriminating between regimes of small, $|F|/F_c\lesssim 1$, and strong compression, $|F|/F_c\gtrsim 1$.
Similarly, the regime of large compression (and stretching) forces can be measured with respect to the characteristic length $\sqrt{\kappa/|F|}$. In particular, we find that $\sqrt{|F|/F_c}\propto L/\sqrt{\kappa/|F|}$ and thus $L\gtrsim \sqrt{\kappa/|F|}$ corresponds to strong compression (and stretching) forces~\cite{Kierfeld:2004,Benetatos:2010}.

To derive an approximate solution of the Fokker-Planck equation [Eq.~\eqref{eq_fokker_planck_3d}]  valid for small temperatures, we rely on an Eikonal approximation~\cite{kleinert:2009}, where
the partition sum is presented in terms of the Gibbs free energy, $Z=\exp(-G/k_\text{B}T)$.
Inserting the partition sum into Eq.~\eqref{eq_fokker_planck_3d} and neglecting higher order terms in the inverse temperature yields
\begin{align}
 \partial_s G\!-\!|F|\cos\vartheta \ G+\frac{1}{2\kappa}\left[\!\left(\partial_\vartheta G\right)^2\!+\!\frac{1}{\sin^2\vartheta}\left(\partial_\varphi G\right)^2\!\right]=0,
\end{align}
where the compressive force is written as $F= -|F|$. We derive the corresponding Euler-Lagrange equations for the angles by applying the method of characteristics,
\begin{align}
 \kappa \frac{\diff^2\vartheta(s)}{\diff s^2}-\frac{\kappa}{2} \sin2\vartheta(s)\left(\!\frac{\diff\varphi(s)}{\diff s}\!\right)^2\!\!
 +|F|\sin\vartheta(s)&=0,\label{eq:euler_lagrange}\\
\frac{\diff}{\diff s}\left(\left[\sin\vartheta(s)\right]^2\frac{\diff\varphi(s)}{\diff s}\right)&=0.
\end{align}
These can also be obtained by minimizing the total energy [Eq.~\eqref{eq:hamiltonian}]~\cite{Pilyugina:2017}. Taking into account the boundary conditions $\diff\vartheta(L)/\diff s = \diff\vartheta(0)/\diff s =0$ and that bending only occurs in a plane at fixed $\varphi$, Eq.~\eqref{eq:euler_lagrange} reduces to
\begin{align}
 \kappa \frac{\diff^2\vartheta(s)}{\diff s^2}+|F|\sin\vartheta(s)&=0,
\end{align}
in agreement with the Euler-Lagrange equation of a rod in 2D~\cite{Kurzthaler:2017}. Then, the end-to-end distance projected along the applied force has been elaborated~\cite{Landau:1986,Baczynski:2007},
\begin{align}
\langle X\rangle &= \sqrt{\frac{2\kappa}{|F|}}\int_0^{\varphi_{L/2}}\diff \varphi \ \frac{\cos\varphi}{\sqrt{\cos\varphi-\cos\varphi_{L/2}}},
\end{align}
for compression forces $|F|>F_c$.

\section{Results -- elastic behavior}
We elucidate the elastic behavior of a semiflexible polymer under compression in terms of the experimentally accessible force-extension relations and the associated susceptibilities and consider the experimental setups of clamped, cantilevered, and free polymers. We compare our results for polymers in 3D with theoretical predictions for polymers confined to 2D and work out the dimensional differences. Moreover, we provide exact solutions for the force-extension relations of polymers subject to pulling forces.

\subsection{Force-extension relation and susceptibility}
\begin{figure*}[htp]
 \includegraphics[width = \linewidth, keepaspectratio]{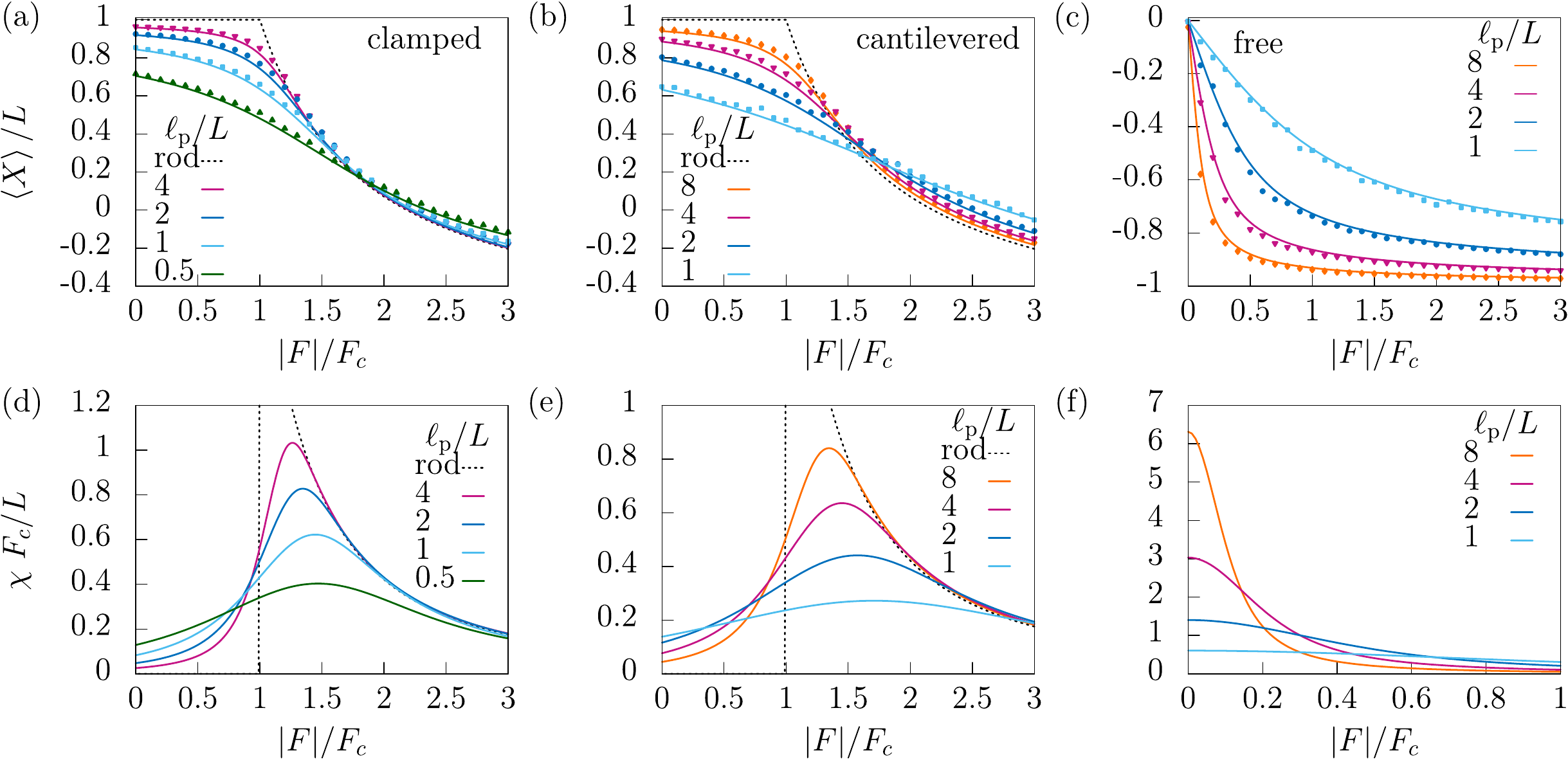}
\caption{Force-extension relation $\langle X\rangle$ projected along the applied compression force $\vec{F}=-|F|\vec{e}$ (\textit{top row})
and associated susceptibility $\chi$ (\textit{bottom row}) for clamped ((a) and (d)), cantilevered ((b) and (e)), and free semiflexible polymers ((c) and (f)) with different persistence lengths $\ell_\text{p}$. Here, $L$ denotes the contour length of the polymer and the critical Euler buckling force $F_c=\pi^2\kappa/(\gamma L)^2$ is used to normalize the forces (with $\gamma=1$ for clamped and $\gamma=2$ for cantilevered and free polymers). The black dashed lines in panels (a), (b), (d), and (e) represent the force-extension relation and the susceptibility of a rigid rod. Symbols correspond to pseudo-Brownian simulations (see appendix~\ref{appendix_simulations}).
\label{fig:compression}}
\end{figure*}
We obtain the force-extension relation of a semiflexible polymer by numerical differentiation of the exact Gibbs free energy [Eq.~\eqref{eq:X}] (see appendix~\ref{appendix_numerics}) and corroborate our solutions for selected parameters by stochastic simulations (see appendix~\ref{appendix_simulations}).

\textit{Clamped and cantilevered polymers}. Our results predict a smooth crossover from an almost stretched to a buckled configuration with increasing compression force (see Figs.~\ref{fig:compression}~(a) and (b)). These findings are in striking contrast to the classical Euler buckling instability of a stiff rod, which is fully stretched at small compression forces and yields for the first time at the critical Euler buckling force $F_c$~\cite{Landau:1986}. Even in the force-free case semiflexible polymers do not display a completely straight configuration, as a rigid rod, due to the presence of thermal fluctuations. In particular, their mean projected end-to-end distance is shorter than their contour length $\langle X\rangle/L<1$ and approaches zero for flexible polymers, $\ell_\text{p}/L\ll 1$. In fact, cantilevered polymers display a mean projected end-to-end distance comparable to that of clamped polymers with approximately half the persistence length.

Interestingly, in the vicinity of the critical Euler buckling force $F_c$, the force-extension relations of semiflexible polymers of different persistence lengths intersect, which reveals a stiffening of the polymers due to thermal fluctuations. Thus, flexible polymers oppose more strongly the applied compression force, whereas they bend more at small forces. For forces exceeding the critical Euler buckling force, $|F|/F_c\gtrsim1$, the mean projected end-to-end distance becomes negative reflecting the alignment of the polymer along the direction of the compression force. In particular, for these forces the force-extension relation of rather stiff polymers, $\ell_\text{p}/L\gtrsim 1$, approaches closely the behavior of a rigid rod.

Comparison of the force-extension relation of a clamped to a cantilevered polymer reveals, that cantilevered polymers respond more sensitively to the applied force, as the critical Euler buckling force is four times smaller ($\gamma^2=4$) than that of a clamped polymer. In fact, the free end of a cantilevered polymer can freely align along the applied force contrary to a clamped end, which can therefore resist the compression more strongly.

\begin{figure*}[htp]
 \includegraphics[width = \linewidth, keepaspectratio]{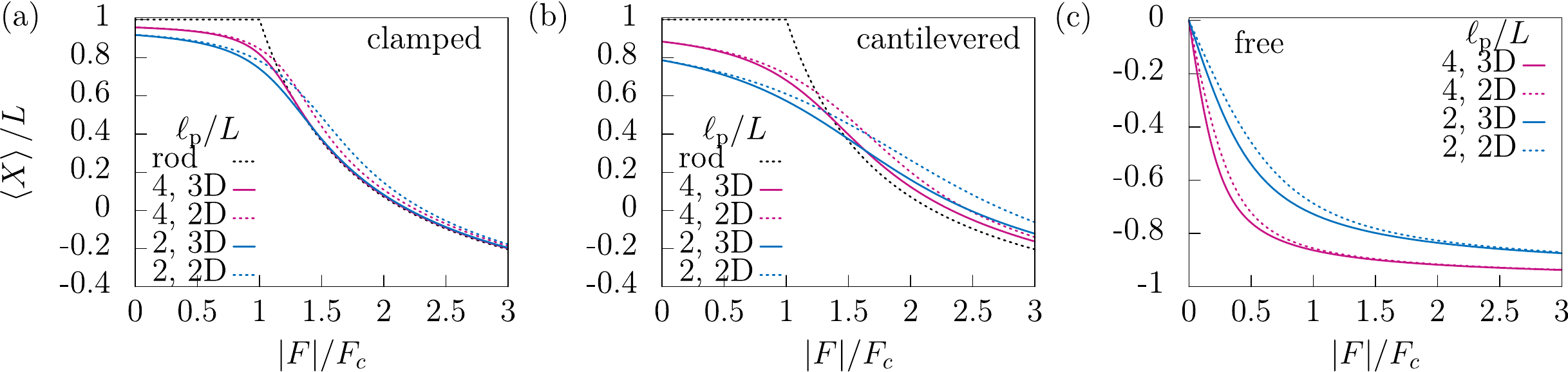}
\caption{Comparison of the force-extension relation $\langle X \rangle$ of clamped, cantilevered, and free
semiflexible polymers in 3D (solid lines) and 2D (dashed lines).
Here, $L$ denotes the contour length and $\ell_\text{p}$ the persistence length of the polymer. The applied compression force
 $|F|$ is normalized by the critical Euler buckling force $F_c$.
\label{fig:comparison}}
\end{figure*}

As a measure to characterize the strength of the response of the polymer to the applied compression force, we have evaluated the susceptibilities $\chi$ of clamped and cantilevered polymers (see Figs.~\ref{fig:compression} (d) and (e)). In both cases, the susceptibilities display a prominent maximum in the vicinity of the critical Euler buckling force indicating, via the fluctuation-response theorem, that the fluctuations of the force-extension relation are most important here. We refer to this regime as the \textit{buckling transition} of the semiflexible polymer, reminiscent of the Euler buckling instability of a stiff rod. In particular, the maximal susceptibility is located at forces $|F|/F_c\gtrsim 1$, which approach the critical Euler buckling force for increasing persistence lengths. Moreover, the peaks become more narrow for polymers with increasing persistence length and come closer to the susceptibility of a stiff rod at forces $|F|/F_c\gtrsim 1$. For comparison, the susceptibility of a stiff rod vanishes for forces smaller than $F_c$, assumes  $\chi(F_c+0) = 2L/F_c$ directly at the transition, and decreases monotonically for increasing forces.

\textit{Free polymers}. In striking contrast to clamped and cantilevered polymers, free polymers exhibit a
qualitatively different response to compression forces (see Fig.~\ref{fig:compression}~(c)). In particular, the force-extension relation remains negative at all forces, which reflects that the ends of the polymer are free to align along the direction of the applied force and the polymer in fact experiences a stretching force rather than compression. At large forces, the mean projected end-to-end distance approaches almost the contour length of the polymer, yet thermal fluctuations prevent the polymer from assuming a fully stretched configuration. Thus, more flexible polymers can resist the applied force more strongly than their stiffer counterparts. The associated susceptibility displays a maximum at vanishing forces independent of the persistence length of the polymer and decreases monotonically for increasing forces (see Fig.~\ref{fig:compression}~(f)).

\textit{Pseudo-dynamic simulations}. We have corroborated our theoretical predictions for the force-extension relations of semiflexible polymers with different boundary conditions by stochastic simulations, see  symbols in Fig.~\ref{fig:compression} (a)-(c). For details on the simulations we refer to the appendix~\ref{appendix_simulations}. Overall we find good agreement between the theory and the simulations, apart from a small systematic error. This can be traced back to the discretization of the contour of the semiflexible polymer, so that in the simulations the polymer appears slightly stiffer due to the finite number of segments. Refining the discretization and simulating over longer times should diminish this discrepancy.

\subsection{Comparison to the force-extension relation of a semiflexible polymer in 2D}
Here, we elucidate the quantitative differences between the force-extension relation obtained for semiflexible polymers in 3D and the behavior of polymers confined to 2D~\cite{Kurzthaler:2017} (see Fig.~\ref{fig:comparison}).
We observe that for vanishing forces the mean projected end-to-end distances for clamped and cantilevered polymers in 2D and in 3D coincide, as anticipated by the definition of the persistence length $\ell_\text{p}$. For instance, for cantilevered polymers
it evaluates to
\begin{align}
\begin{split}
 \langle X\rangle_{F=0} &\!=\! \int_0^L\!\diff s \ \left\langle\vec{u}_0\cdot\vec{u}(s)\right\rangle\!=\! \ell_\text{p}\left[1-\exp(-L/\ell_\text{p})\right],
\end{split}
\end{align}
independent of the dimension. For forces in the vicinity of the critical Euler buckling force $|F|/F_c\gtrsim 1$, polymers in 3D yield more strongly than their confined counterparts, which reflects that the confinement in fact helps the polymer to resist the applied compression force. Interestingly, the force-extension relations of polymers in 2D approach those of polymers in 3D for large forces indicating that dimensional differences are indeed most prominent at the buckling transition, where the susceptibilities are maximal. Moreover, differences are more pronounced for cantilevered polymers than for clamped polymers.

For polymers with free ends the mean projected end-to-end distance vanishes at zero force due to rotational symmetry (see Fig.~\ref{fig:comparison}~(c)). At small forces differences become apparent between the 2D and 3D case, which show that polymers confined to 2D can oppose the applied force more strongly, than those in 3D. However, both force-extension relations agree for increasing forces and thus our results predict that the behavior of the polymer with respect to the dimension of the system is most sensitive at small forces, $|F|/F_c\lesssim 1$.

\subsection{Semiflexible polymer under tension}
\begin{figure}[htp]
\centering
 \includegraphics[width = \linewidth, keepaspectratio]{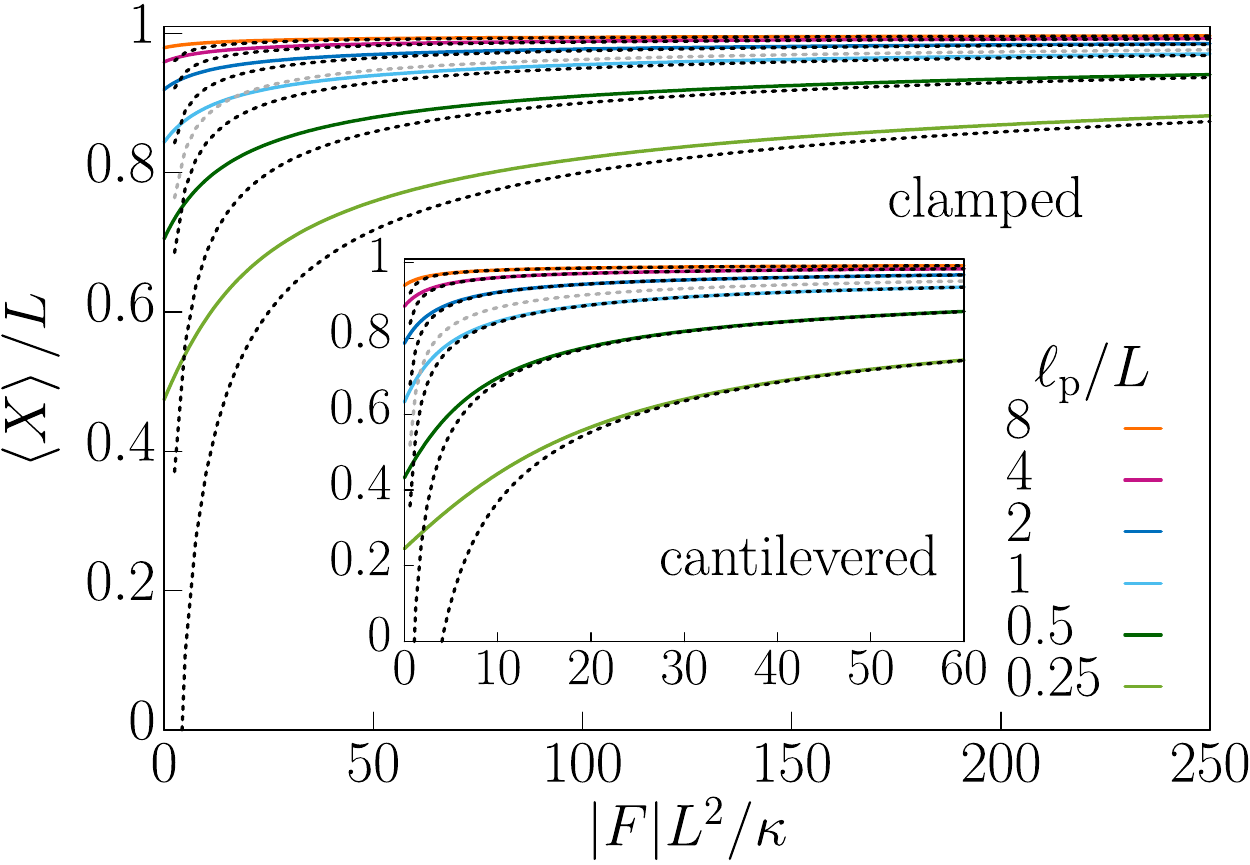}
\caption{Force-extension relation $\langle X \rangle$ of clamped and cantilevered (\textit{inset}) semiflexible polymers under tension $\vec{F}=|F|\vec{e}$. Here, $L$ denotes the contour length and $\ell_\text{p}$ the persistence length of the polymer, which is related to the bending rigidity $\kappa$ via the thermal energy, $\ell_\text{p}=\kappa/k_\text{B}T$. The black dashed lines correspond to the large-force asymptote [Eq.~\eqref{eq:wb}]~\cite{Marko:1995}. The gray dashed lines indicate the large-force approximation from Ref.~\cite{Ha:1997}.
\label{fig:tension}}
\end{figure}
The force-extension relations of a polymer with free ends under tension are up to a sign identical to those under compression and have already been presented in Fig.~\ref{fig:compression}~(c). In contrast, the elastic behavior of clamped and cantilevered polymers under tension is qualitatively different to their buckling behavior as the clamped end(s) introduce a preferred direction of alignment (see Fig.~\ref{fig:tension}). In particular, we have elaborated for the first time exact solutions for the corresponding force-extension relations, which  increase monotonically with the applied force and thus reflect the stretching of the polymer at increasing pulling force. Yet, thermal fluctuations remain important, particularly for flexible polymers, where strong forces are required to extend the polymer to almost its contour length. Due to its free end, a cantilevered polymer can oppose the applied force more strongly and the force-extension relation is always smaller than that of a clamped polymer with the same persistence length. Furthermore,  our analytic solutions of cantilevered polymers approach the well-known approximation of Marko and Siggia~\cite{Marko:1995},
\begin{align}
\frac{\langle X\rangle}{L} = 1-\frac{1}{\sqrt{4\ell_\text{p}|F|/k_\text{B}T}}, \label{eq:wb}
\end{align}
in the regime of stiff polymers $\ell_\text{p}/L\gg 1$ and for large forces $|F|L^2/\kappa\gg 1$ (corresponding to $L\gg \sqrt{\kappa/|F|}$). Yet, for more flexible polymers at small forces the approximate theory deviates from our exact results, as angular fluctuations become relevant and the weakly-bending assumption is no longer fulfilled.

Interestingly, we find that the approximate theory [Eq.~\eqref{eq:wb}] also approaches the stretching behavior of clamped polymers for increasing stiffness and forces $|F|L^2/\kappa\gg 1$, irrespective of the boundary conditions, as has been predicted by Ref.~\cite{Benetatos:2017}. Yet, stronger forces are required than for cantilevered polymers.

There have been other approaches to calculate force-extension relations of stretched semiflexible polymers. These rely, for example, on treating the local inextensibility constraint $|\vec{u}(s)|=1$ within a mean-field theory~\cite{Ha:1997}. The resulting large force asymptote $\langle X\rangle/L = 1-3/4\sqrt{\ell_{p}|F|/k_\text{B}T}$ obeys the same scaling as Eq.~\eqref{eq:wb} but with a different prefactor. We find that for $\ell_\text{p}/L=2$ (gray dashed lines in Fig.~\ref{fig:tension}) it approximates the analytic solution (similarly for $\ell_\text{p}/L=4,8$ (not shown)), yet deviations are still visible for the full parameter range considered, which increase for more flexible polymers. In addition to the mean-field approach for an inextensible polymer, Ref.~\cite{Blundell:2009} has also considered an extensible polymer and predicted mean elongations exceeding the contour length of the polymer due to the mechanical stretching.

\section{Summary and conclusion}
We have obtained exact expressions for the partition sum of a wormlike chain in 3D subject to external forces in terms of generalized spheroidal wave functions. Derivatives of the corresponding exact Gibbs free energy have provided the force-extension relations and the associated susceptibilities of clamped, cantilevered, and free semiflexible polymers of arbitrary stiffnesses. In the case of clamped and cantilevered polymers under compression the force-extension relation decreases monotonically with the applied load, from an almost stretched configuration at vanishing force to a buckled configuration at strong compression.  The force-extension relations of semiflexible polymers of different persistence lengths intersect, revealing a stiffening of the polymers due to thermal fluctuations. Interestingly, the associated susceptibilities exhibit a prominent maximum at forces in the vicinity of the critical Euler buckling force of a stiff rod, which indicates, via the fluctuation-response theorem, that the fluctuations of the force-extension relations become most prominent here. Moreover, the forces at the maximal susceptibility represent the analog to the critical Euler buckling force of a stiff rod for semiflexible polymers. We have compared our analytical predictions for a polymer in 3D to the behavior of a polymer confined to 2D~\cite{Kurzthaler:2017} and find that confinement favors the resistance of the polymer to the compression force. Free polymers display a qualitatively different elastic behavior, as both ends are free to align with the applied force and the polymers essentially experience a pulling force.

Previous predictions for the buckling behavior of semiflexible polymers in 3D have been provided in Ref.~\cite{Pilyugina:2017}, where the energy due to compression is related to the given end-to-end distance, $\mathcal{H}_\text{force} = |F||\vec{r}(L)-\vec{r}(0)|$, but leaves the ends of the polymer free. Here, the exact force-extension relation of the semiflexible polymer has been extracted from the minimum of the Helmholtz free energy which has been obtained from the associated Green function. In this work the Green function of a semiflexible polymer has been calculated exactly via a continued fraction approach. Similar to our results for clamped and cantilevered polymers, the force-extension relations for polymers of different persistence lengths intersect in the vicinity of the critical Euler buckling force.

Yet, Monte Carlo simulations show that the corresponding configurations display a U-shape, where the ends of the polymer approach each other for increasing forces. These features differ from our analysis where at least one end is clamped along the direction of the force and the stretching energy depends on the alignment along the direction of the force [Eq.~\eqref{eq:H_force}] rather than on the distance between both ends. For the 2D analog of our model, clamped polymers assume an S-shaped and cantilevered polymers a hooked-shaped configuration at large compression forces~\cite{Kurzthaler:2018}, where the clamped ends remain aligned along the direction of the force (i.e. opposite to the compression). Interestingly, at the \textit{buckling transition} the associated probability density becomes bimodal reflecting that semiflexible polymers can either resist the applied force and display an almost elongated configuration or bend. Similarly, the appearance of almost stretched and U-shaped configurations has also been found in the vicinity of the critical Euler buckling force by Ref.~\cite{Pilyugina:2017}. In this scenario the force always acts compressive and thus the ends of the polymers come close to each other at large compression forces so that the force-extension relations approach zero. This behavior is in striking contrast to our model, where the polymer can orient along the force leading to a negative projected end-to-end distance. In particular, free polymers display a negative force-extension relation at all forces, as they can flip and experience stretching forces rather than compression.

To gain a profound understanding about the configurational properties of semiflexible polymers in 3D under compression, future studies should elucidate the probability densities for the projected end-to-end distance. In particular, analytical predictions can be achieved by formally replacing the force by a complex force $|F|\mapsto |F|+\imath k/k_\text{B}T$, where $k$ denotes the wavenumber of the corresponding characteristic function. A Fourier transform of the characteristic function then yields the probability density.

Furthermore, our solution strategy for the partition sum of a semiflexible polymer has exploited that the ensemble of its configurations is identical to the ensemble of trajectories of a single self-propelled agent. In particular, the Fokker-Planck equation, which describes the evolution of the partition sum of a semiflexible polymer as the contour length $s$ increases, can be mapped to the equation of motion for the characteristic function of the random displacements of an active Brownian particle~\cite{Kurzthaler:2016,Kurzthaler:2018:janus}. Thus, we anticipate that the elastic properties of a semiflexible polymer with a spontaneous curvature can be worked out by using the mathematical analog of a Brownian circle swimmer~\cite{Kurzthaler:2017:circle}.

Our analytic predictions for semiflexible polymers in 3D provide the basis for experimentally measured force-extension relations of semiflexible polymers and permit to extract key properties, such as the persistence length of the polymers. Yet, for the analysis of the elastic properties of topologically confined biopolymers, such as cytoskeletal polymers or DNA strands in cells, in \emph{vitro} experiments have been concerned with polymers immersed in, for example, biaxially confined micro-channels~\cite{Choi:2005} or quasi-2D chambers~\cite{Witz:2011}. Here, the force-extension relations for 2D semiflexible polymers~\cite{Kurzthaler:2017} could be used to capture the spatial confinement.

The response of a semiflexible polymer to a force applied to its end serves as reference to understand the more intricate behavior of polymers suspended in external fields, including, for example, electric~\cite{Ferree:2003,Li:2012} and force fields~\cite{Lamura:2001,Benetatos:2004} or hydrodynamic flows~\cite{Perkins:1995,Larson:1997,Wandersman:2010}. It also constitutes a convenient input to elucidate the impact of hinge defects present along the contour of specific biopolymers, such as double-stranded DNA strands~\cite{Padinhateeri:2013,Benetatos:2017}.
Moreover, the elasticity of a single polymer serves as ingredient to understand the complex properties of polymer networks, which are formed by entangled semiflexible polymers~\cite{Bausch:2006, MacKintosh:2014}. In particular, our results provide a theoretical footing for the stiffening of cross-linked networks at small compression~\cite{Claessens:2006,Chaudhuri:2007}, where the constituents of the network in fact experience pulling forces and stretch laterally. Yet, in the vicinity of the critical
Euler buckling force of the single components a reversible softening of these networks occurs, since the filaments aligned with the direction of the applied force cannot oppose it anymore and start to buckle~\cite{Chaudhuri:2007}. These nonlinear elastic features have also been found in computer simulations of cross-linked networks~\cite{Amuasi:2015}, where our analytic results could serve as accurate input parameter to model the response of the single components upon confinement.

Furthermore, the elasticity of a single polymer in free space, as elaborated here, could be used as a reference to study the behavior of a single polymer in crowded environments composed of fixed obstacles~\cite{Hofling:PRL:2008,Hofling:2008,Schobl:2014}, solutions of  polymers~\cite{Leitmann:2016, Leitmann:2017, Lang:2018}, or active constituents~\cite{Schaller:2010,Kaiser:2014,Shelley:2016}, as, for example, molecular motors. These give rise to intriguing non-equilibrium physics resembling transport processes inside cells and providing fundamental input for novel active materials~\cite{Ramaswamy:2010}.

\section*{Conflicts of interest}
There are no conflicts to declare.

\section*{Acknowledgments}
I particularly thank Thomas Franosch for very helpful and stimulating discussions and a critical reading of the manuscript. This work has been supported by the Austrian Science Fund (FWF) via the contract No. P~28687-~N27.

\begin{widetext}
\section*{Appendix}
\appendix
\section{Numerical evaluation of the generalized spheroidal wave functions\label{appendix_numerics}}
The generalized spheroidal wave functions $\text{Ps}_{\ell}^m(R,\eta)$ are expanded in terms of
the associated Legendre polynomials [Eq.~\eqref{eq:Pslm}]. For the evaluation of the partition sums only
the generalized spheroidal wave functions of order $m=0$ are required, which
reduce to an expansion in the Legendre polynomials
$|j,m=0\rangle\equiv|j\rangle:=P_j(\eta)\sqrt{(2j+1)/2}$ for $j\in \mathbb{N}_0$,
\begin{align}
  \text{Ps}_{\ell}^0(R,\eta) 	&= \sum_{j=0}^\infty d^{0\ell}_j(R) |j\rangle,\label{eq:Pslndl}
\end{align}
with corresponding orthonormalization for the coefficients $\sum_{j=0}^\infty d_j^{0\ell}(R)d_j^{0k}(R)=\delta_{\ell k}$.
We use a bra-ket notation to emphasize the analogy to methods familiar from
quantum mechanics and the associated scalar product corresponds to an integral,
$\langle j|\ell\rangle=\sqrt{(2j+1)(2\ell+1)}\int_{-1}^1\!\diff\eta \ P_j(\eta)P_\ell(\eta)/2$ for $j,\ell\in \mathbb{N}_0$.
Moreover, integration over the generalized spheroidal wave functions yields
\begin{align}
 \int_{-1}^1\! \diff \eta \ \text{Ps}_\ell^0(R,\eta) &= \sum_{j=0}^\infty d^{0\ell}_j(R) \sqrt{\frac{2j+1}{2}}\int_{-1}^1\!\diff \eta \ P_j(\eta)=\sqrt{2} \ d_0^{0\ell}(R). \label{eq:int_Ps}
\end{align}

Inserting the expanded eigenfunction [Eq.~\eqref{eq:Pslndl}] into the eigenvalue problem [Eq.~\eqref{eq:spheroidal_EV}]
and projecting onto $\langle n|$ leads to the matrix eigenvalue problem
\begin{align}
\sum_{j}\left[\left\langle n|R\eta|j\right\rangle -n(n+1)\delta_{jn}\right]d_j^{0\ell}(R) &= -A^0_{\ell}(R)d_n^{0\ell}(R). \label{eq:ABP_matrix}
\end{align}
Since the matrix elements are non-vanishing for $j=n-1,n,n+1$ only,
it is in fact a tridiagonal matrix.
In particular, we evaluate the integrals
$\langle n| \eta|\ell\rangle 	= \sqrt{(2n+1)(2\ell+1)}\int_{-1}^1 \diff \eta \ \text{P}_n(\eta) \eta\text{P}_\ell(\eta)/2$
by using the properties of the Legendre polynomials. The matrix eigenvalue problem simplifies to
\begin{align}
\alpha_n d^{0\ell}_{n+1}(R)-\left[n(n+1)-A_{\ell}^0(R)\right] d^{0\ell}_n(R)+\beta_n d^{0\ell}_{n-1}(R)&=0, \label{eq:recurrence_di}
\end{align}
with coefficients
\begin{align}
\alpha_n	&= R\frac{n+1}{\sqrt{(2n+1)(2n+3)}} \qquad \text{ and } \qquad \beta_n	= R \frac{n}{\sqrt{(2n+1)(2n-1)}}.
\end{align}
The normalized eigenvectors $\vec{d}^{0\ell}(R)\equiv\vec{d}^{0\ell}= \left[d_0^{0\ell},d_1^{0\ell},d_2^{0\ell},...\right]^{\text{T}}$
and eigenvalues $A^0_{\ell}(R)\equiv A^0_{\ell}$ can be efficiently determined numerically by solving the corresponding matrix eigenvalue problem.
We truncate the matrix in Eq.~\eqref{eq:ABP_matrix} to sufficiently high order (here and throughout $\dim = 100$)
such that the normalization of the lowest eigenmode, $\sum_{\ell} \left[d_0^{0\ell}\right]^2=1$, is achieved.
In practice we sort the eigenvalues with increasing size, $A_0^0\leq A_1^0\leq\dots$,
which induces a natural cutoff of the series in Eqs.~\eqref{eq:SumZ_clamped}-\eqref{eq:SumZ3}.

We have used the multiprecision library \textit{mpmath} in python~\cite{mpmath} for the numerical evaluation of the generalized spheroidal wave functions, the computation of the partition sums [Eqs.~\eqref{eq:SumZ_clamped}-\eqref{eq:SumZ3}], and the derivatives of the corresponding Gibbs free energy to obtain our results for the force-extension relations [Eq.~\eqref{eq:X}] and susceptibilities [Eq.~\eqref{eq:chi}].

\section{Pseudo-dynamic simulations\label{appendix_simulations}}
To corroborate our analytic predictions, we perform stochastic simulations of the
semiflexible polymer under compression. Here, we derive an algorithm for Brownian dynamics
simulations of a discretized semiflexible polymer, which reproduce the canonical ensemble
as stationary state, similar to our previous work on polymers in 2D~\cite{Kurzthaler:2017}.
Since our aim is merely to validate our analytic results for the equilibrium properties of
a semiflexible polymer, we neglect features such as hydrodynamic interactions,
anisotropic friction, or the translational motion of the polymer.
In particular, an equation of motion for the probability density needs to be formulated with the requirement
that its stationary distribution concurs with the equilibrium distribution. Standard methods of
stochastic calculus can then be applied to obtain
the corresponding stochastic differential equations from the Fokker-Planck equation of the probability density~\cite{Gardiner:2009}.

We first discretize the contour $L$ of the polymer equidistantly in terms of the positions of
the beads $\{\vec{R}_i\}_{i=0}^{N}$. The corresponding tangent vectors $\{\vec{u}_i\}_{i=0}^{N-1}$
are obtained by $\vec{u}_i=(\vec{R}_{i+1}-\vec{R}_i)N/L$ and have unit length $|\vec{u}_i|=1$.
Thus, discretization of the Hamiltonian in Eq.~\eqref{eq:hamiltonian} yields
\begin{align}
    \frac{\mathcal{H}(\{\vec{u}_i\}_{i=0}^{N-1})}{k_\text{B}T}&=
    \frac{\hat{\ell}_\text{p}}{2}\sum_{i=0}^{N-2}(\vec{u}_{i+1}-\vec{u}_{i})^2-\hat{f}\sum_{i=0}^{N-1}\vec{e}\cdot\vec{u}_i,\
\end{align}
where  $\hat{\ell}_\text{p}=\ell_\text{p}N/L$ is the scaled persistence length and
$\hat{f}=fL/N$ denotes the scaled force with $\hat{f}>0$ tension
and $\hat{f}<0$ compression forces, respectively.
As the tangent vectors fulfill the inextensibility constraint, we parametrize them in terms of the spherical
coordinates $\{\vartheta_i,\varphi_i\}\equiv\{\vartheta_i,\varphi_i\}_{i=0}^{N-1}$,
with polar angles $\vartheta_i=\angle (\vec{u}_i,\vec{e})$ measured relative to the direction $\vec{e}$ of the applied force
and azimuth angles $\varphi_i$. Thus, the Hamiltonian evaluates to
\begin{align}
\begin{split}
            &\frac{\mathcal{H}(\{\vartheta_i,\varphi_i\})}{k_\text{B}T}=
            \hat{\ell}_p\sum_{i=0}^{N-2}[1-\cos\vartheta_{i+1}\cos\vartheta_i-\cos(\varphi_{i+1}-\varphi_i)\sin\vartheta_{i+1}\sin\vartheta_i]
             -\hat{f}\sum_{i=0}^{N-1}\cos\vartheta_i. \label{eq:H_dis}
\end{split}
\end{align}
The associated probability density of the polymer in equilibrium then reads
\begin{align}
            \mathbb{P}_\text{eq}(\{\vartheta_i,\varphi_i\}) &= Z^{-1} \exp\left[-\frac{\mathcal{H}(\{\vartheta_i,\varphi_i\})}{k_\text{B}T}\right]\sqrt{|g|}, \label{eq:P_eq}
\end{align}
with determinant of the metric $g=\prod_i\sin^2\vartheta_i$. The probability density fulfills the normalization
$\int \left[\prod_{i=0}^{N-1} \diff \varphi_i \diff\vartheta_i \right]  \mathbb{P}_\text{eq}(\{\vartheta_i,\varphi_i\})=1$.

Next, we derive the Fokker-Planck equation for the conditional probability density
$\mathbb{P}\equiv\mathbb{P}(\{\vartheta_i,\varphi_i\},t|\{\vartheta_i^0,\varphi_i^0\})$, that a
polymer with initial orientations $\{\vartheta_i^0,\varphi_i^0\}$ at time $t=0$ has changed its orientations to
$\{\vartheta_i,\varphi_i\}$ at time $t$.
Thus, the time evolution of the conditional probability density is expressed as
\begin{align}
  \partial_t \mathbb{P} = -\sum_{i=0}^{N-1}\sum_{q_i = \vartheta_i,\varphi_i}\partial_{q_i} \left[U^{q_i}(\{\vartheta_i, \varphi_i\})\mathbb{P}\right], \label{eq_FP_time}
\end{align}
where $U^{q_i}(\{\vartheta_i,\varphi_i\})$ denotes the velocity of the probability current for the coordinate $q_i$.  It is obtained by the friction law
$U^{q_i}(\{\vartheta_i,\varphi_i\})=\sum_k K^{q_iq_k}(\{\vartheta_i,\varphi_i\})F_{q_k}(\{\vartheta_i,\varphi_i\})$ with
rotational mobility tensor $K^{q_iq_k} = N\xi_r^{-1}\delta_{ik}$ and forces,
\begin{align}
  F_{q_k} &=-\partial_{q_k}\mathcal{H}-k_\text{B}T\partial_{q_k}\ln\mathbb{P}+k_\text{B}T\partial_{q_k}\ln\sqrt{|g|}. \label{eq:forces}
\end{align}
Here, the first term corresponds to the mechanical forces, the second accounts for the thermal Brownian forces, and the third term represents the metric forces due to the inextensibility constraints of the segments, $|\vec{u}_i|=1$. In particular, by construction, the equation of motion reproduces the equilibrium distribution [Eq.~\eqref{eq:P_eq}] in the stationary state, $\partial_t\mathbb{P}=0$.
Collecting our results, the Fokker-Planck equation for the conditional probability density evaluates to
\begin{align}
\begin{split}
&\partial_t \mathbb{P}=
  \hat{D}_\text{rot}\sum_{i=0}^{N-1}\Bigl\{\partial_{\vartheta_i}\Bigl[\hat{\ell}_\text{p}\bigl[\sin\vartheta_i(\cos\vartheta_{i+1}+\cos\vartheta_{i-1})
-\cos\vartheta_i\bigl(\cos(\varphi_{i+1}-\varphi_{i})\sin\vartheta_{i+1}
 +\cos(\varphi_{i}-\varphi_{i-1})\sin\vartheta_{i-1}\bigr)\bigr]\\
&\ \ +\hat{f}\sin\vartheta_i-\cot\vartheta_i\Bigr]\mathbb{P}
+\hat{\ell}_\text{p}\partial_{\varphi_i}\bigl[\sin\vartheta_{i}\bigl(\sin(\varphi_{i}-\varphi_{i-1})\sin\vartheta_{i-1}
 -\sin(\varphi_{i+1}-\varphi_{i})\sin\vartheta_{i+1}\bigr)\bigr]\mathbb{P}+\partial^2_{\vartheta_i}\mathbb{P}+\partial^2_{\varphi_i}\mathbb{P}\Bigr\},\label{eq:FPsim}
\end{split}
\end{align}
where $\hat{D}_\text{rot}=Nk_\text{B}T/\xi_\text{r}$
denotes the scaled rotational diffusion coefficient.
The initial condition reads
\begin{align}
\begin{split}
  \mathbb{P}(&\{\vartheta_i,\varphi_i\},t=0|\{\vartheta_i^0,\varphi_i^0\},0) =
\prod_{i=0}^{N-1}\delta(\vartheta_i-\vartheta_i^0\text{ mod }\pi)\delta(\varphi_i-\varphi_i^0\text{ mod }2\pi).
\end{split}
\end{align}
We employ standard methods of stochastic calculus~\cite{Gardiner:2009} to derive the Langevin
equations for the angles, $\vartheta_i(t)\equiv\vartheta_i$ and $\varphi_i(t)\equiv\varphi_i$, from the Fokker-Planck equation [Eq.~\eqref{eq:FPsim}]
\begin{align}
\begin{split}
\diff\vartheta_i =& - \hat{D}_\text{rot}\Bigl[\hat{\ell}_\text{p}\bigl[\sin\vartheta_i(\cos\vartheta_{i+1}+\cos\vartheta_{i-1})
-\cos\vartheta_i\bigl(\cos(\varphi_{i+1}-\varphi_{i})\sin\vartheta_{i+1}
 -\cos(\varphi_{i}-\varphi_{i-1})\sin\vartheta_{i-1}\bigr)\bigr]\\
&\ \ \ \ \ \ \ \ \ +\hat{f}\sin\vartheta_i-\cot\vartheta_i\Bigr]\diff t+\sqrt{2\hat{D}_\text{rot}}\omega_i(t)\diff t,
\end{split}\\
\begin{split}
\diff\varphi_i  =& - \hat{D}_\text{rot}\hat{\ell}_\text{p}\sin\vartheta_{i}\bigl(\sin(\varphi_{i}-\varphi_{i-1})\sin\vartheta_{i-1}
-\sin(\varphi_{i+1}-\varphi_{i})\sin\vartheta_{i+1}\bigr)\diff t
+\sqrt{2\hat{D}_\text{rot}}\zeta_i(t)\diff t,
\end{split}
\end{align}
where $\omega_i (t)$ and $\zeta_i(t)$ are independent Gaussian white noise processes with zero mean $\langle \omega_i(t)\rangle=\langle \zeta_i(t)\rangle =0$ and delta-correlated variance $\langle \omega_i(t) \omega_j(t')\rangle=\langle \zeta_i(t) \zeta_j(t')\rangle = \delta(t-t')\delta_{ij}$ for $i,j = 1,\hdots,N-2$. These equations of motion encode the pseudo-dynamics for the discretized polymer chain. We adapt the initial and final angles, $\vartheta_0, \varphi_0$ and $\vartheta_{N-1}, \varphi_{N-1}$, (and their time evolution) according to the boundary conditions.

To obtain reliable statistics we have performed $10$ realizations of a polymer with $N=100$ segments and a time step of $10^{-5}/\hat{D}_\text{rot}$ over a time horizon of $10^4/\hat{D}_\text{rot}$. Configurations of the polymer are measured after it has equilibrated.
\end{widetext}

\bibliography{polymer_bib}

\end{document}